\documentclass[preprint,authoryear,3p,times,12pt]{interact}
\usepackage[utf8]{inputenc}
\usepackage[table,xcdraw]{xcolor}
\usepackage{amsfonts,amsmath,bm,bbm}
\usepackage{graphicx}
\usepackage{epstopdf}
\usepackage{amssymb}
\usepackage{mathtools}
\usepackage{enumerate}
\usepackage{natbib}
\usepackage{indentfirst}
\usepackage[font=footnotesize]{caption}
\usepackage{floatrow}
\usepackage{enumitem}
\DeclareMathOperator{\argmin}{argmin}
\usepackage{subcaption}
\usepackage{url}
\usepackage{tikz}
\usepackage{varwidth}
\usetikzlibrary{patterns,decorations.pathreplacing}
\usepackage{booktabs}
\usepackage[normalem]{ulem}
\useunder{\uline}{\ul}{}
\usepackage{multirow}

\usepackage{color}
\definecolor{gray}{rgb}{0.5,0.5,0.5}
\definecolor{dgreen}{rgb}{0,0.5,0}

\begin{document}
		
		\title{Smoothing Quantile Regression Averaging: A new approach to probabilistic forecasting of electricity prices}
		
		\author{
			\name{Bartosz Uniejewski\thanks{Email: bartosz.uniejewski@pwr.edu.pl, ORCID: 0000-0001-8563-2514}}
			\affil{Department of Operations Research and Business Intelligence, Wroc{\l}aw University of Science and Technology, 50-370 Wroc{\l}aw, Poland}
		}
		
		\maketitle
		
		\begin{abstract}
			Accurate short-term price forecasting is essential for daily operations in electricity markets. This article introduces a new method, called Smoothing Quantile Regression (SQR) Averaging, that improves upon well-performing probabilistic forecasting schemes. To demonstrate its utility, a comprehensive study is conducted on two electricity markets, including recent data covering the COVID-19 pandemic and the Russian invasion of Ukraine. The performance of SQR Averaging is evaluated both in terms of reliability and sharpness measures, and economic benefits from a trading strategy. The latter utilizes battery storage and sets limit orders using selected quantiles of the predictive distribution. SQR Averaging leads to profit increases of up to 3.5\% on average compared to the benchmark strategy based solely on point forecasts. This is strong evidence for the practical value of using probabilistic forecasts in day-ahead power trading, even in the face of the COVID-19 pandemic and geopolitical disruptions.
		\end{abstract}
		

	\section{Introduction}
	
	Recent events such as the COVID-19 pandemic and geopolitical risks in Europe and the Middle East have added instability to an already volatile power market. As a result, accurate short-term price forecasting has become more important than ever before.  With intraday volatility reaching even up to 690 EUR (for German EPEX market), day-ahead predictions have become critical for daily operations in the day-ahead, intraday or balancing markets.
	Accurate day-ahead price forecasts are critical for market participants to make informed decisions about their bidding strategies \citep{nar:zie:22}. If participants have a reliable price forecast, they can adjust their bids accordingly and increase their chances to execute profitable transactions.

	In recent decades, most studies on electricity price forecasting (EPF) have focused primarily on point forecasts. However, the Global Energy Forecasting Competition 2014 (GEFCom2014) marked a shift by advocating the computation of quantile (probabilistic) forecasts instead \citep{hon:pin:fan:etal:16}. In comparison to point predictions, the probabilistic ones provide a full picture of the future price distribution. They allow for a better adjustment of the offering process to avoid the costs induced by a range of unexpected generation or consumption volumes \citep{mor:con:mad:pin:zug:14}. Probabilistic EPF is a concept closely related to risk management. In general, good quality probabilistic price forecasts can help producers, retailers, and speculators to determine optimal strategies for short-term operations \citep{uni:wer:21}, as well as derivative pricing, Value-at-Risk calculations, hedging, and trading \citep{bun:and:che:wes:16}. 
	
	To improve the accuracy of quantile forecasts, this paper introduces a novel approach to obtaining probabilistic forecasts called Smoothing Quantile Regression (SQR) Averaging\footnote{The term \textit{quantile regression averaging} is used here to refer to the general concept of combining quantile regression with forecast averaging techniques. However, it is also used specifically to refer to the Quantile Regression Averaging model introduced by \citet{now:wer:15}. To avoid confusion, the general concept will be referred to as \textit{QR Averaging}, while the model proposed by \citet{now:wer:15} as \textit{QRA}. Similarly, \textit{SQR Averaging} will refer to the idea of using the smoothing quantile regression technique to average forecasts, while \textit{SQRA} to a specific model.}. 
	Like Quantile Regression (QR) Averaging utilizes quantile regression \citep{koe:05} to average point predictions, SQR Averaging uses the recently introduced smoothing quantile regression \citep{fer:gue:hor:21}. Three variants are considered: SQRA -- a smoothed equivalent of Quantile Regression Averaging \citep[QRA;][]{now:wer:15}, SQRM -- a smoothed equivalent of Quantile Regression Machine \citep[QRM;][]{mar:uni:wer:20}, and SQRF -- a smoothed equivalent of Quantile Regression with probability (F) averaging \citep[QRF;][]{uni:mar:wer:19}.
	
	Model performance is evaluated using datasets from two major European electricity markets, from Germany and Spain. Empirical results demonstrate the superior predictive performance of the new method compared to well-established benchmarks in terms of the \citet{kup:95} test, the pinball score \citep{gne:kat:14}, as well as the conditional predictive accuracy test \citep{gia:whi:06}.   
	
	While traditional evaluation methods focus on statistical error measures, the assessment of economic value is paramount \citep{yar:pet:21}. Few studies have quantified the economic benefits of electricity price forecasting, and a standardized valuation methodology has yet to be established \citep{hon:pin:etal:20,mac:uni:wer:22}. Recent approaches, such as that of \cite{uni:wer:21}, introduce trading strategies that use probabilistic predictions to evaluate the performance of forecasting models.
	
	This paper proposes a new trading strategy that can be applied in day-ahead markets by market participants, having access to an energy storage system. Such systems have become more prevalent due to the growth of renewable energy production \citep{gia:rav:ros:20}.  The proposed trading strategy is inspired by that of \citet{uni:wer:21}, but here each transaction is made in the day-ahead market without the need to use the balancing market to close the position.
	
	The remainder of this paper is organized as follows.  Section \ref{sec:data} provides a brief description of the datasets. Section \ref{sec:methodology} discusses the construction of point forecasts, introduces the idea of smoothing quantile regression, along with a description of the probabilistic forecasting benchmarks, and presents the trading strategy. In Section \ref{sec:results}, the predictive performance of point forecasts is evaluated, followed by a discussion of the coverage and sharpness of probabilistic forecasts. In addition, they are evaluated in terms of the profits obtained using the proposed trading strategy. Finally, Section \ref{sec:conclusion} summarizes the results and draws conclusions.
	
	\section{Datasets}
	\label{sec:data}
	
		\begin{figure}[t!]
		\centering
		\includegraphics[width=0.95\textwidth]{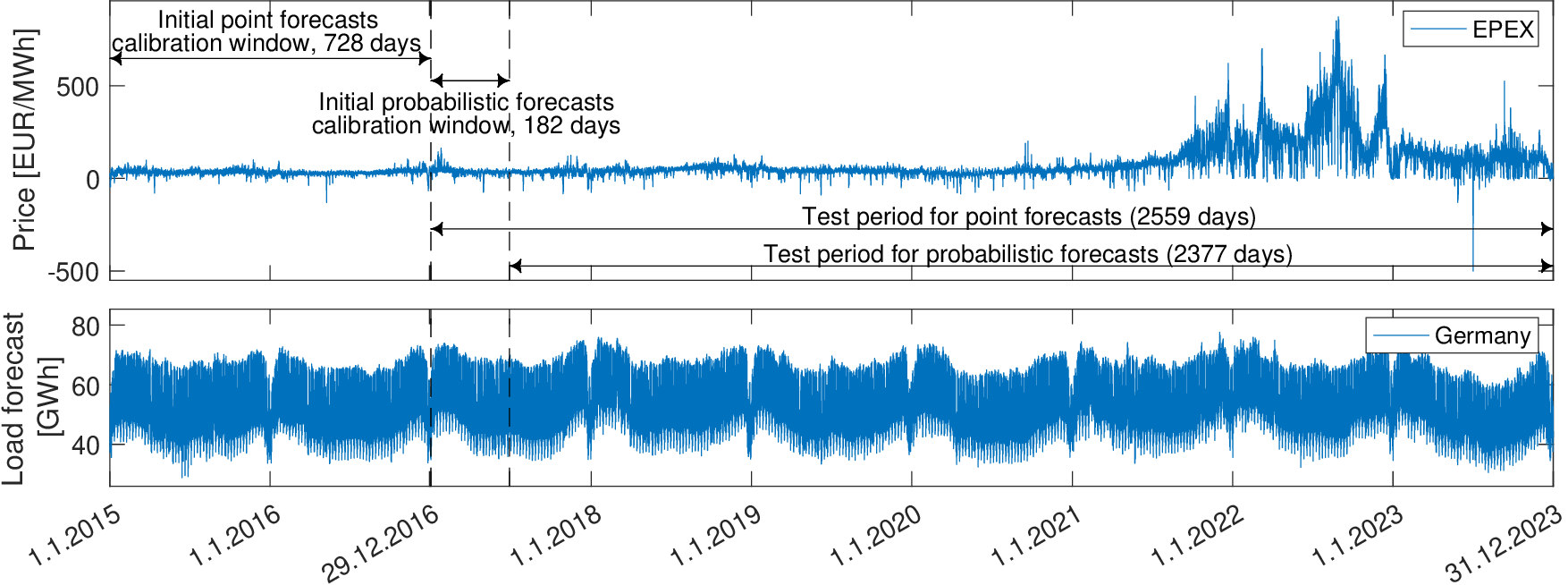}
		
		\includegraphics[width=0.95\textwidth]{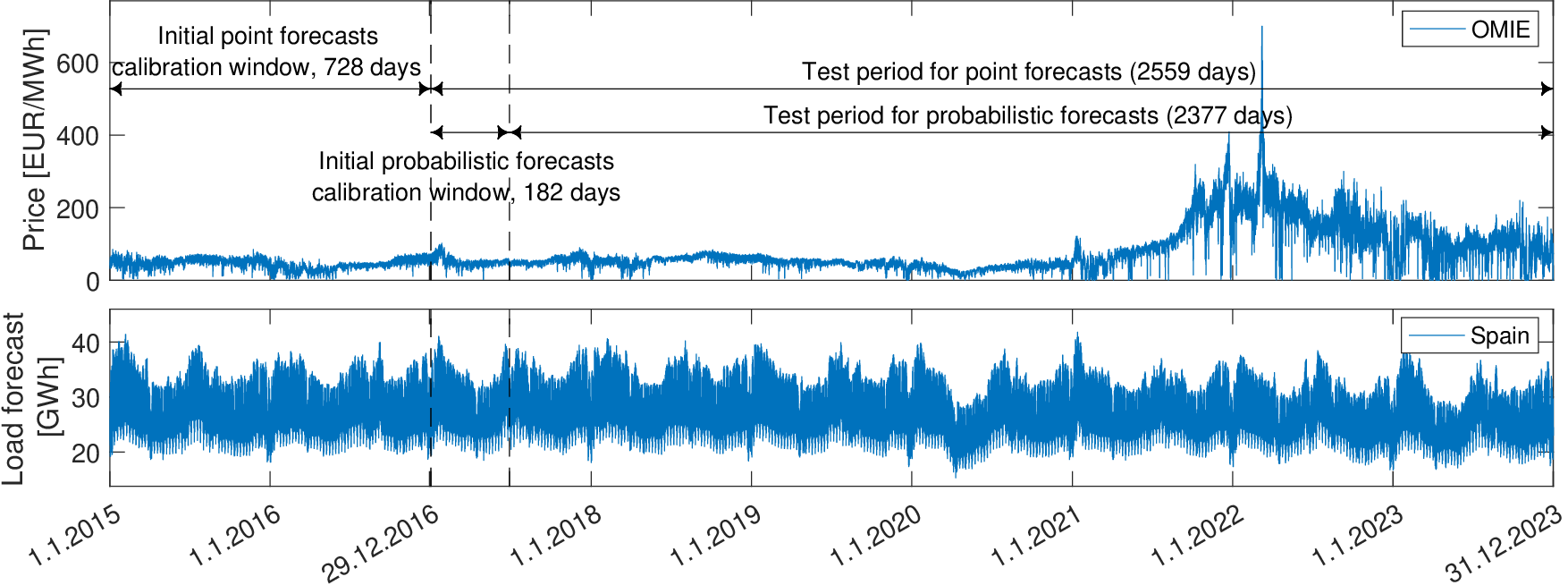}
		
		\caption{Day-ahead prices and day-ahead load forecasts for the German EPEX (top) and Spanish OMIE (bottom) markets from 1.1.2015 to 31.12.2023. The vertical dashed lines mark respectively the beginning of the out-of-sample test period for point forecasts (29.12.2016; also the beginning of the initial 182-day calibration window for probabilistic forecasts) and the beginning of the out-of-sample test period for probabilistic forecasts (29.6.2017). The first 728 days constitute the initial calibration window for point forecasts.}
		\label{fig:data1}
	\end{figure}

	This empirical study use datasets from two European markets, each spanning nine years. The first, the German EPEX spot market, has recently experienced a large increase in the share of renewable generation and some major structural changes \citep{kap:let:wei:23}. The data was collected from the ENTSO-E transparency platform (\url{https://transparency.entsoe.eu}) and consists of two time series, each with hourly resolution (see the top panel of Figure \ref{fig:data1}):
	\begin{itemize} \itemsep 0pt
		\setlength{\parskip}{0pt}
		\item day-ahead electricity prices for the DE-AT-LU bid zone until 30.9.2018 and DE-LU afterwards,
		\item day-ahead total load forecasts for Germany.
	\end{itemize}
	Due to its central location in Europe, interesting price dynamics with negative prices and high trading liquidity in most segments, the German market is one of the most studied \citep[for example][]{hag:etal:16,cal:fus:ron:17,kat:zie:18,mac:22,mar:nar:wer:zie:23,nit:wer:23}. 
	
	The second market is the Spanish OMIE market. This market is also very often analyzed in the literature \citep[for example][]{bar:etal:09,gar:car:san:15,cia:mar:nas:23,lip:uni:wer:24,uni:24}. It is interesting for analysis because of the rapid increase in the share of renewable energy, for example, the installed capacity of solar power generation increased from 6500 MW in 2015 to 18500 MW in 2023.
	The data have been collected from the transparency platform and include two time series with an hourly resolution (see the bottom panel of Figure \ref{fig:data1}): 
	\begin{itemize}\itemsep 0pt
		\setlength{\parskip}{0pt}
		\item day-ahead marginal electricity prices for Spain,
		\item day-ahead total load forecasts for Spain.
	\end{itemize}
	It had a very narrow admissible price range of [0, 180] EUR, until the second half of 2021, then it was changed to [-500 and 3000] EUR.

	All time series cover 3287 days from 1.1.2015 to 31.12.2023. Missing or duplicated values (corresponding to the time changes) were replaced by the average of the closest observations for missing hours and the arithmetic mean of the two values for duplicated hours.
	
	
	The selection of these two datasets from major European energy markets serves to underscore the robustness of the proposed method and its applicability to different time series. In particular, both datasets exhibit significant changes in the price series structure over the nine-year time period, as illustrated in Figure \ref{fig:data1}. The onset of the COVID-19 pandemic in 2020, followed by the Russian invasion of Ukraine in 2022, led to significant shifts in electricity price dynamics. 
	
	\section{Methodology}
	\label{sec:methodology}
	
	Point forecasts are determined using a rolling calibration window scheme, using data from the last 728 days (104 weeks or about 2 years). First, the models described in the Section \ref{ssec:point} are calibrated using data from 1.1.2015 to 28.12.2016, and forecasts are determined for all 24 hours of 29.12.2019. Then the window is moved forward one day (i.e. 1.1.2015 is removed and 29.12.2016 is added) and the forecasts for all 24 hours of 30.12.2016 are computed. This procedure is repeated until the forecasts for the last day in the dataset (31.12.2023) are made, giving a total of 2559 forecast days.

	Once the point forecasts are obtained, they are used as a basis to construct 	probabilistic forecasts. All of the approaches considered (see Section \ref{ssec:Probabilistic}) require a subsample of the point forecasts. Therefore, a rolling calibration window of 182 days (26 weeks) is used to estimate the quantiles of the future price distribution. As a result, probabilistic forecasts are obtained and then evaluated over the 2377-day period from 29.6.2017 to 31.12.2023 (see Figure~\ref{fig:data1}).
	
	The remainder of this section describes the forecasting scheme, which consists of two stages. First, Section \ref{ssec:point} introduces the point forecasting models. The Section \ref{ssec:Probabilistic} presents probabilistic forecasting benchmarks and introduces the smoothing quantile regression averaging models. Finally, Section \ref{ssec:strategies} proposes and describes a probabilistic forecast-based trading strategy.
	
	\subsection{Point forecasts}
	\label{ssec:point}
	
	For each day in the out-of-sample test period for point forecasts, i.e., $d=1,2,...,2377$, for each hour of the day, i.e., $h=1,2,...,24$, and for each of the considered variance stabilizing transformations (VSTs), the same multi-step procedure is followed:
	\begin{equation}
		P_{d,h} \xrightarrow{\text{standardize}} p_{d,h} \xrightarrow{\text{VST}} Y_{d,h} \xrightarrow{\text{predict}} \hat{Y}_{d,h} \xrightarrow{\text{inverse VST}} \hat{p}_{d,h}. \xrightarrow{\text{inverse standardization}} \hat{P}_{d,h}.
	\end{equation}
	First, the prices in the 728-day calibration window are standardized. Next, one of the transformation functions described in the Section \ref{ssec:VST} is applied. Then, the parameters of the expert model, see Eq.\ \eqref{eqn:expert}, are estimated using ordinary least squares and day-ahead point forecasts are computed. Finally, the inverse of the transformation used in the second step and the inverse of the standardization are applied to obtain the final price forecasts. The same standardization and VST are applied to the load forecast series prior to fitting the expert model in the forecasting step. As a result of this routine, for each day $d$ and hour~$h$, five different point forecasts of the same expert model (see Section \ref{ssec:expert}) estimated with different VSTs are obtained.
	
	\subsubsection{Variance Stabilizing Transformations}
	\label{ssec:VST}
	
	Electricity price spikes are usually caused by unpredictable weather conditions, power outages, or transmission failures \citep{gia:gro:12}. These, in turn, can significantly bias electricity price forecasts \citep{gro:nan:19}. The outliers pull the model coefficients toward values that better fit the spikes and increase the in-sample errors for the non-spiky periods. The general idea of variance stabilizing transformations is to reduce the variation in the data \citep{cia:mun:zar:22}. A lower variation or less spiky behavior of the input data usually allows the forecasting model to produce more accurate predictions \citep{jan:tru:wer:wol:13}. 

Following \cite{uni:wer:zie:18}, before applying a proper VST, each time series (prices and load) is first standardized by subtracting the sample median ($a$) and dividing by the sample mean absolute deviation ($b$). Finally, the transformation is applied and the transformed price is denoted by $Y_{d,h}$, i.e. $Y_{d,h} = f(p_{d,h})$, where~$f(\cdot)$ is a given VST. After computing the forecasts, the inverse transformation and standardization are applied to obtain the final electricity price forecasts: $\hat{P}_{d,h} = bf^{-1}(\hat{Y}_{d,h})+a$.

In this study, motivated by the results of \citet{uni:wer:zie:18}, five VST classes were used: \emph{area hyperbolic sine} (asinh), \emph{Box-Cox} (boxcox), \emph{mirror log} (mlog), \emph{polynomial} (poly) and \emph{normal distribution probability integral transform} (N-PIT). A detailed description of each transformation can be found in the original paper \citep{uni:wer:zie:18}. 

\subsubsection{The expert model}
\label{ssec:expert}

Due to the computational complexity required to produce probabilistic forecasts within a rolling calibration window scheme, the computational effort required to obtain point forecasts must be limited. Therefore, sophisticated techniques 
are replaced by a parsimonious autoregressive structure. The model for the transformed price at day $d$ and hour $h$ is given by \citep{zie:wer:18}:
\begin{eqnarray}
	\label{eqn:expert}
	Y_{d,h} & = & \underbrace{\beta_1 Y_{d-1,h} + \beta_2 Y_{d-2,h} + \beta_3 Y_{d-7,h}}_{\mbox{\scriptsize autoregressive effects}} + \underbrace{\beta_4 Y_{d-1,24}}_{\mbox{\scriptsize end-of-day}} + \underbrace{\beta_5 Y_{d-1}^{max} + \beta_6 Y_{d-1}^{min}}_{\mbox{\scriptsize non-linear effects}} \nonumber \\
	&& + \underbrace{\beta_7 L_{d,h}}_{\mbox{\scriptsize load}} + \underbrace{\sum\nolimits_{j=1}^7 \beta_{h,j+7} D_{j}}_{\text{weekday dummies}} + \varepsilon_{d,h}, 
\end{eqnarray}
where $Y_{d-1,h}$, $Y_{d-2,h}$ and $Y_{d-7,h}$ represent the autoregressive terms and refer to the prices at the same hour of the previous day, two days earlier, and one week earlier, respectively. $Y_{d-1,24}$ is the midnight price of electricity for the previous day and is the last known price at the time of the forecast. $Y^{max}_{d-1}$ and $Y^{min}_{d-1}$ are the maximum and minimum prices of the previous day. $L_{d,h}$ is the transformed day-ahead load forecast for a given hour of a day, finally $D_{1},..., D_{7}$ are weekday dummies, and $\varepsilon_{d,h}$ is the noise term, assumed to be independent and identically distributed variables.

\subsection{Probabilistic forecasts}
\label{ssec:Probabilistic}

A common way to compute probabilistic forecasts is to construct quantile forecasts. Instead of predicting the expected value of the future price, we predict the price range within which the future price will fall with a given probability ($\alpha$). If we continue with this idea and extend it to construct multiple prediction intervals (PIs), the end result will be a set of quantiles of many different levels. In this research, the goal is to predict 99 percentiles ($q = 1\%,2\%,...,99\%$), which is a reasonably accurate approximation of the full distribution. 


\subsubsection{Johnson distribution (ARX-J)}
\label{sssec:ARX-J}

The first benchmark is based on the assumption that the electricity price follows the \citet{joh:49} distribution (JSU). \citet{gia:bun:17} compare the goodness-of-fit to electricity data of several distribution including the normal (Gaussian) and JSU distributions and conclude that the latter performs significantly better. Recently the JSU distribution was used in the context of EPF by \cite{mar:nar:wer:zie:23}. The model is defined as:

\begin{equation}
	\hat P^{q}_{d,h} = \hat{P}_{d,h}+\gamma^{\text{J}}_{q}, 
\end{equation}
where $\hat{P}_{d,h}$ is the point forecast and $\gamma_{q}$ is the $q$ quantile of the Johnson's distribution of the point forecast errors ($\hat{\varepsilon}_{d,h}$) with parameters estimated in the probabilistic calibration period. The model uses the average of five point forecasts (each corresponding to different VSTs) to derive $\hat{\varepsilon}_{d,h}$.

\subsubsection{Historical simulation (HS)}
\label{sssec:hist}

Historical simulation, directly constructs probabilistic forecasts by considering both point forecasts and forecast errors. The quantile forecast ($\hat P^{q}_{d,h}$) is defined as

\begin{equation}
	\hat P^{q}_{d,h} = \hat{P}_{d,h}+\gamma^{\text{HS}}_{q}, 
\end{equation}
where $\gamma_{q}$ is the $q$ quantile of $\hat{\varepsilon}_{d,h}$. As before, the model uses the ensemble of five individual point forecasts. .

\subsubsection{Conformal Prediction (CP)}
\label{sssec:cp}

Conformal Prediction computes prediction intervals (PI) based on absolute point forecast errors within a calibration window \citep{sha:vov:08}. The prediction intervals are centered on the point forecast, which limits the generation of quantile forecasts unless the errors are assumed to be symmetrically distributed. This paper adopts the approach of \cite{kat:zie:21}, where
\begin{equation}
	\hat P^{q}_{d,h} = 	
	\begin{cases}
		\hat{P}_{d,h}-\gamma^{\text{CP}}_{q}, & \mbox{for } q<50\%,\\
		\hat{P}_{d,h}+\gamma^{\text{CP}}_{q}, & \mbox{for } q>50\%],
	\end{cases} 
\end{equation}
and the parameter $\gamma^{\text{CP}}_{q}$ is derived as the $|1-2q|$th quantile of $|\hat{\varepsilon}_{d,h}|$.

\subsubsection{Quantile Regression (QR) Averaging}
\label{sssec:QRA}

Quantile regression averaging was introduced by \citet{now:wer:15} and is widely used in the electricity price forecasting literature \citep{jan:mic:20, kat:zie:21,mac:uni:wer:22}. QR Averaging is a special case of quantile regression, where the independent variables (regressors) are point forecasts of the dependent variable, i.e. ${\hat P}^{(i)}_{d,h}$ for $i=1,...,n$. To obtain parameter estimates, the sum of the so-called check functions is minimized \citep{koe:05}: 

\begin{equation}\label{eq:qr:loss}
	\hat{\boldsymbol\beta}_q
	= \underset{\boldsymbol\beta_{q}}\argmin \Big\{ \textstyle\sum_{d,h} \underbrace{\left({q}-\mathbbm{1}_{{P}_{d,h}< {X_{d,h}}\boldsymbol\beta_{q}}\right)\left({P}_{d,h} - X_{d,h} \boldsymbol\beta_q\right)}_{\mbox{\scriptsize \emph{check function}}} \Big\}.
\end{equation}

Despite its superb performance in GEFCom2014, the method can have problems when the set of regressors is larger than a few \citet{mar:uni:wer:20}. In addition, the obtained interval predictions tend to be too narrow, making the electricity price forecast less reliable. To address these problems, \citet{wan:etal:19} proposed adding a constraint to the model so that the parameters are non-negative and sum to one. More recently, \citet{uni:wer:21} suggested using $\ell^1$ regularization to correctly select the inputs for QRA. While this solution yields significantly lower pinball loss, it does not completely solve the problem of too narrow prediction intervals. 

\subsubsection{Smoothing Quantile Regression (SQR) Averaging}
\label{sssec:SQRA}

In a recent article, \citet{fer:gue:hor:21}, propose to modify the standard quantile regression by smoothing the objective function using kernel estimation. They show that the proposed solution not only yields a lower mean squared error compared to the standard QR estimator, but is also asymptotically differentiable. In addition, the asymptotic theory holds uniformly with respect to quantile level and bandwidth.

Here, it is proposed to used the smoothing quantile regression of \citet{fer:gue:hor:21} to average a pool of point forecasts, creating a novel technique to obtain probabilistic forecast directly from point predictions. The estimator of $\boldsymbol\beta_{q}$ takes following form:
\begin{equation}\label{eq:sqr:loss}
	\hat{\boldsymbol\beta}_q
	= \underset{\boldsymbol\beta_{q}}\argmin \Big\{ \textstyle\sum_{d,h}  H \times \phi\left(\frac{{P}_{d,h} - X_{d,h} \boldsymbol\beta_q}{H}\right) + \left(q - \Phi\left(\frac{{P}_{d,h} - X_{d,h} \boldsymbol\beta_q}{H}\right)\right) \left({P}_{d,h} - X_{d,h} \boldsymbol\beta_q \right)  \Big\},
\end{equation}
where the independent variable matrix $X_{d,h}$ is the matrix of $n$ point forecasts ${\hat P}^{(i)}_{d,h}$ for $i=1,\ldots, n$, $\Phi(\cdot)$ and $\phi(\cdot)$ are the cumulative distribution function (cdf) and the probability density function (pdf) of the standard normal distribution, respectively, while $H$ is the bandwidth.

\begin{figure}[t]
	\centering
	\includegraphics[width=.95\textwidth]{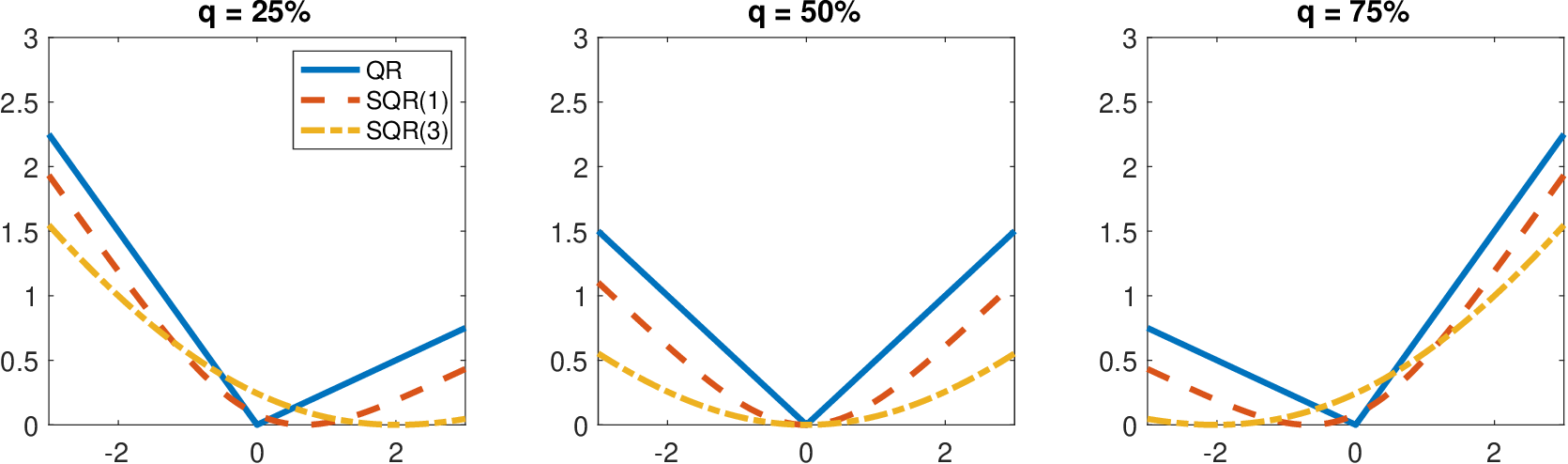}
	\caption{Comparison of the shape of the objective function for the  quantile regression and smoothing quantile regression estimators across quantile levels. Note that the SQR objective function has been shifted downward for clarity, as its minimum does not reach the X-axis.}
	\label{fig:qr_vs_sqr}
\end{figure}

To better illustrate the difference between smoothing and standard QR, the objective functions for different quantile levels are plotted in Figure \ref{fig:qr_vs_sqr}. It can be observed that for any quantile (except the median $q=50\%$), the minimum of the SQR objective function is not centered around 0. Consequently, the favorable situation for predicting one of the lower quantiles is when the prediction is less than the actual price. Conversely, the lowest value of the objective function for the upper quantiles is shifted to the left of the Y-axis. This means that the SQR expects the upper quantile to be above the actual price and gives a higher penalty when the upper quantile prediction hits the actual price, which is the perfect scenario for the standard QR. The size of the shift, as well as the slope of the objective function, depends on the bandwidth $H$. Note that the SQR is actually a generalization of the standard QR. The limit of the objective function of the SQR, when $H$ tends to 0, is equivalent to the standard quantile regression.

When comparing the prediction intervals (PIs) obtained with the QR and SQR estimators, the latter are usually significantly wider. Recall that probabilistic forecasts of electricity prices most often suffer from PIs that are too narrow, which is why the SQR solution seems to fit the electricity data perfectly.

\subsubsection{Bandwidth selection}

As with many applications of kernel estimation, the performance of smoothing quantile regression depends strongly on the choice of the bandwidth. Following the recommendation of \citet{fer:gue:hor:21}, a simple approach is used to select the optimal bandwidth. The bandwidth value is calculated according to the Rule-of-Thumb (RoT) method of \citet{sco:92}:

\begin{equation}\label{eq:sco}
	H = 1.06\frac{\sigma_\varepsilon}{N^{\frac{1}{5}}},
\end{equation}
where $\sigma_\varepsilon = \min \lbrace \text{std}, \text{IQR} \rbrace$ is a minimum of the standard deviation and interquartile range of the in-sample residuals of standard quantile regression, while $N$ is the number of observations in the sample (here $N=182$).

\subsubsection{QR and SQR Averaging based models}
\label{sssec:models}

\begin{figure}[t]
	\tikzset{%
		every neuron/.style={
			circle,
			draw=gray,
			minimum size=1.3cm,
			fill = white,
			text = black
		},
		every neuronprob/.style={
			circle,
			draw=gray!30,
			minimum size=1.3cm,
			fill = gray!30,
			text = black
		},
		our brace/.style={
			thick,	
			decoration={brace,raise=-0.2cm}, 
			decorate,
		},
		our brace down/.style={
			thick,	
			decoration={brace,raise=-0.2cm,mirror}, 
			decorate,
		},
	}
	\centering
	\scalebox{1.05}{
		\begin{tikzpicture}[x=1cm, y=0.8cm, >=stealth]
			
			\foreach \l [count=\y] in {(S)QRF}
			\node [every neuronprob/.try, neuron \l/.try, draw = black, text=black]
			(ave2-2) at (-4.5,-3) {\tiny{\l}};
			
			\node [every neuronprob/.try, neuron 1/.try, draw = black]
			(qr-1) at (-1.5,3-1*2) {\begin{varwidth}{0.8cm}\centering\tiny{(S)QR asinh} \end{varwidth}};
			
			\node [every neuronprob/.try, neuron 2/.try, draw = black]
			(qr-2) at (-1.5,3-2*2) {\begin{varwidth}{0.8cm}\centering\tiny{(S)QR boxcox} \end{varwidth}};
			
			\node [every neuronprob/.try, neuron 3/.try, draw = black]
			(qr-3) at (-1.5,3-3*2) {\begin{varwidth}{0.8cm}\centering\tiny{(S)QR mlog} \end{varwidth}};
			
			\node [every neuronprob/.try, neuron 4/.try, draw = black]
			(qr-4) at (-1.5,3-4*2) {\begin{varwidth}{0.8cm}\centering\tiny{(S)QR poly} \end{varwidth}};
			
			\node [every neuronprob/.try, neuron 5/.try, draw = black]
			(qr-5) at (-1.5,3-5*2) {\begin{varwidth}{0.8cm}\centering\tiny{(S)QR N-PIT} \end{varwidth}};
			
			\node [every neuron/.try, neuron 1/.try] 
			(scann-1) at (1,3-1*2) {\begin{varwidth}{1cm}\centering\tiny{Point forecast asinh} \end{varwidth}};
			
			\node [every neuron/.try, neuron 2/.try] 
			(scann-2) at (1,3-2*2) {\begin{varwidth}{1cm}\centering\tiny{Point forecast boxcox} \end{varwidth}};
			
			\node [every neuron/.try, neuron 3/.try] 
			(scann-3) at (1,3-3*2) {\begin{varwidth}{1cm}\centering\tiny{Point forecast mlog} \end{varwidth}};
			
			\node [every neuron/.try, neuron 4/.try] 
			(scann-4) at (1,3-4*2) {\begin{varwidth}{1cm}\centering\tiny{Point forecast poly} \end{varwidth}};
			
			\node [every neuron/.try, neuron 5/.try] 
			(scann-5) at (1,3-5*2) {\begin{varwidth}{1cm}\centering\tiny{Point forecast N-PIT} \end{varwidth}};
			
			\foreach \m [count=\y] in {1}
			\node [every neuron/.try, neuron \m/.try] 
			(qrm-\m) at (3.5,-3+\y*3) {\begin{varwidth}{1cm}\centering\tiny{Averaged point forecast} \end{varwidth}};
			
			\foreach \l [count=\y] in {(S)QRM}
			\node [every neuronprob/.try, neuron \l/.try, draw = black]
			(ave-1) at (6,-3+\y*3) {\tiny{\l}};
			
			\foreach \l [count=\y] in {(S)QRA}
			\node [every neuronprob/.try, neuron \l/.try, draw = black]
			(ave-2) at (6,-3-\y*3) {\tiny{\l}};

			\foreach \i in {1,...,5}
			\draw [dotted, ->, color= black] (qr-\i) -- (ave2-2);	
			
			\foreach \i in {1,...,5}
			\draw [->, color= black] (scann-\i) -- (qr-\i);
			
			\foreach \i in {1,...,5}
			\draw [dashed, ->, color= gray] (scann-\i) -- (qrm-1);
			
			\draw [->, color= black] (qrm-1) -- (ave-1);
			
			\foreach \i in {1,...,5}
			\draw [->, color= black] (scann-\i) -- (ave-2);
			\node [align=center, above] at (-3.25,2.25) {\begin{varwidth}{1.7cm}\centering\scriptsize{Probability averaging} \end{varwidth}};
			\draw [our brace/.try] 
			(-5+0.2,2.25) -- (-1.5-0.2,2.25);
			\node [align=center, above] at (-0.25,2.25) {\begin{varwidth}{1.7cm}\centering\scriptsize{(S)QR procedure} \end{varwidth}};
			\draw [our brace/.try] 
			(-1.5+0.2,2.25) -- (1-0.2,2.25);
			\node [align=center, above] at (2.25,2.25) {\begin{varwidth}{1.7cm}\centering\scriptsize{Point forecast averaging} \end{varwidth}};
			\draw [our brace/.try] 
			(1+0.2,2.25) -- (3.5-0.2,2.25);
			\node [align=center, above] at (4.75,2.25) {\begin{varwidth}{1.7cm}\centering\scriptsize{(S)QR procedure} \end{varwidth}};
			\draw [our brace/.try] 
			(3.5+0.2,2.25) -- (6-0.2,2.25);
			\node [align=center, below] at (3.5,-8.25) {\begin{varwidth}{1.7cm}\centering\scriptsize{(S)QR procedure} \end{varwidth}};
			\draw [our brace down/.try] 
			(1+0.2,-8.25) -- (6-0.2,-8.25);
		\end{tikzpicture}
	}
	\caption{Visualization of the (S)QRA, (S)QRM, and (S)QRF concepts for computing probabilistic forecasts. The point forecasts (corresponding to different VSTs) are represented by white circles, while the gray circles represent probabilistic forecasts. Solid arrows represent the process of constructing probabilistic forecasts from point forecasts, dotted arrows represent the process of averaging probabilistic forecasts, and dashed arrows represent the process of averaging point forecasts.
	}
	\label{fig:ave}
\end{figure}
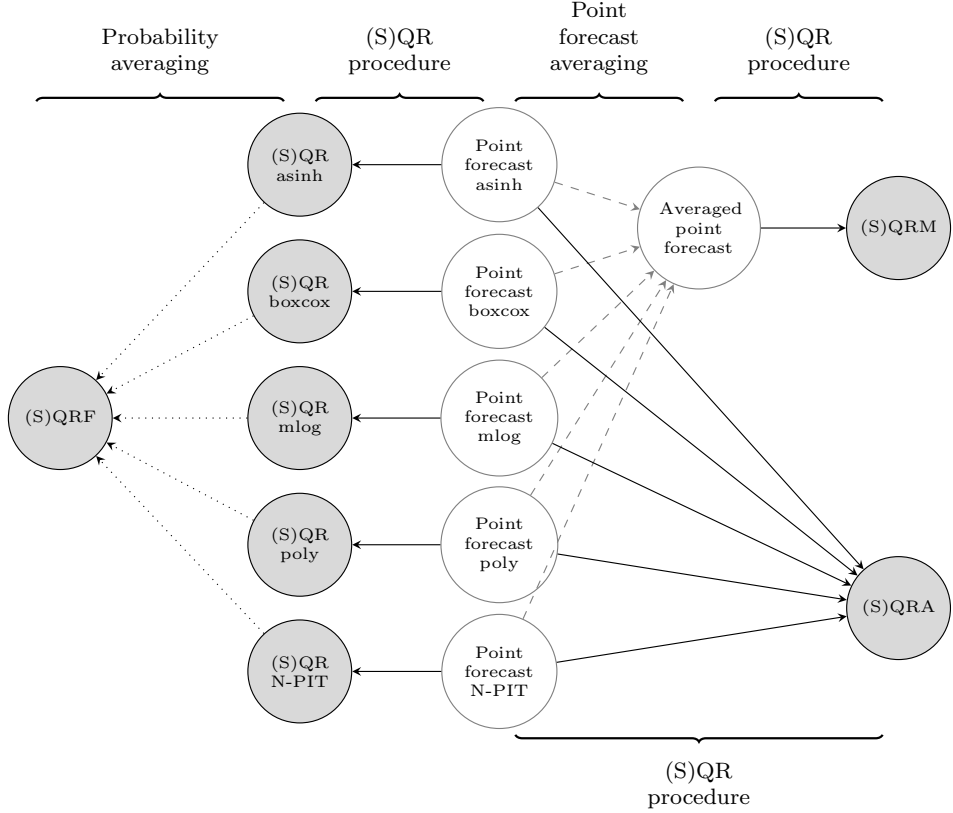

To compute probabilistic forecasts of the day-ahead price, five different point forecasts are used, obtained with one model structure but different data transformation functions (see Section \ref{ssec:point}). Moreover, three variants of averaging are considered in this research (see also Figure \ref{fig:ave}):
\begin{itemize}
	\item The first approach is to apply one of the described methods to all available point forecasts. This solution is referred to as QRA and SQRA for QR and SQR-based models, respectively.
	\item The second approach is to use point forecasts obtained for a single VST to derive a probabilistic forecast corresponding to that VST. As a result, five different probabilistic predictions are obtained, which are then averaged across probabilities \citep[called F or vertical averaging;][]{lic:g-c:win:13}, for details see Section~3.3.2 in the \citet{uni:mar:wer:19}. These models are denoted by QRF and SQRF.
	\item Finally, according to \citet{mar:uni:wer:20}, in some applications it may be beneficial to first average point forecasts and then construct a probabilistic forecast based on fewer but more accurate point predictions. In such models, called QRM or SQRM, all available point forecasts are first combined using the arithmetic mean, and the averaged prediction is used as an input to QR or SQR.
\end{itemize}

To better understand the difference between the described variants, a graph comparing the models is shown in Figure \ref{fig:ave}. Finally, note that the predictions for all 99 percentiles are computed separately. In such a setup, it is possible to obtain a nonmonotonic quantile function. Therefore, after obtaining the predictions of the 99 percentiles, the results are sorted independently for each day and hour. 

\subsection{Trading strategy}
\label{ssec:strategies}

Despite the fact that the primary purpose of forecasting is to support decision making and maximize profits, most EPF studies use only statistical measures to assess forecast accuracy \citep{hon:pin:etal:20,mac:uni:wer:22}, with only a few articles considering the economic value of forecasts. Among them, we can distinguish two basic approaches: the first considers a one-sided (supply or demand) perspective \citep{doo:amj:zar:17,jan:woj:22, mac:22}, and the second evaluates the quality of forecasts based on the trading strategy \citep{uni:wer:21}.

Following the latter approach, this paper proposes a trading strategy for the day-ahead market. It uses probabilistic forecasts to support the decision making process. The strategy is suitable for a company with access to an energy storage system. Without loss of generality, suppose the company owns a 2.5 MWh battery. For both economic and technical reasons, it cannot be discharged below 0.5 MWh (20\% of nominal capacity), and its efficiency is 90\%. The company aims to buy energy and charge the battery when the price is low, and discharge and sell energy when the price is high. In order to trade only in the day-ahead market, the company limits the volume of a single transaction (charge or discharge) to 1~MWh. The proposed strategy aims to leave 1 MWh of energy in the energy storage system at the end of each day. However, depending on whether the submitted bid or offer is accepted or rejected, three battery states are possible at the beginning of the next day:

\begin{itemize}\itemsep 0em
\item $B_d = 0$ refers to the situation when the lower technical limit of the battery is reached at the beginning of day $d$ and there is no more energy to sell. On day $d^*$, when $B_{d^*} = 0$, the company will additionally place a market order (i.e., unlimited price) offer to buy energy and charge the battery.
\item $B_d = 1$ refers to the desirable situation where a day starts with a half-full storage. In this state, the battery can be both charged and discharged. On day $d^*$, when $B_{d^*} = 1$, the battery owner follows the basic steps described in Section \ref{sssec:Quantile-based:strategies}.
\item $B_d = 2$ refers to the situation where the capacity limit has been reached and the battery cannot be charged at the beginning of day $d$. On day $d^*$, when $B_{d^*} = 2$, the company will make an additional unlimited offer to discharge the battery and sell energy.
\end{itemize}

\subsubsection{Quantile-based trading strategies}
\label{sssec:Quantile-based:strategies}

\begin{figure}[tb]
	\centering 
	\includegraphics[width = 0.73 \linewidth]{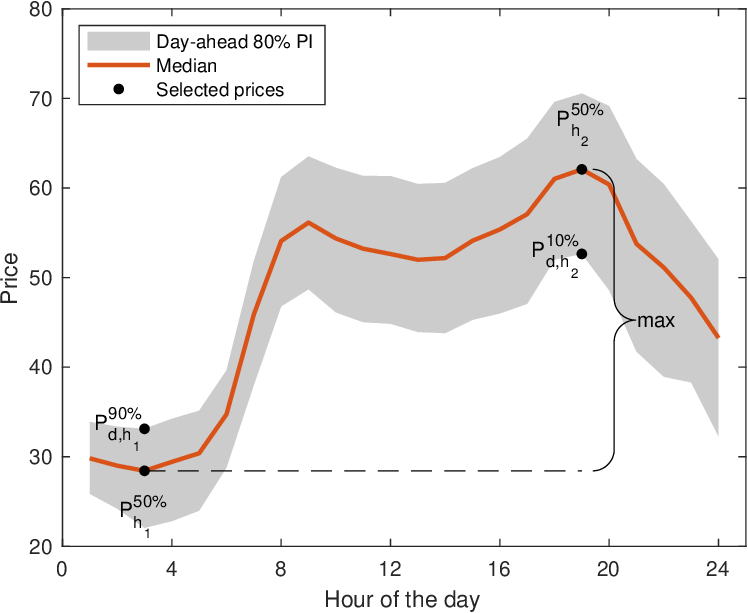}
	\caption{Illustration of the quantile-based bidding strategy for 17.1.2019 for and EPEX market. The orange line shows the median forecast ($\hat P^{50\%}_{d,h}$), the gray area represents the 80\% prediction interval computed one day earlier, and the filled circles at the PI bounds represent the offer price $\hat P^{10\%}_{d,h_2}$ and the bid price $\hat P^{90\%}_{d,h_1}$, see step (2) in Section \ref{sssec:Quantile-based:strategies}.}
	\label{fig:stategy}
\end{figure}

The proposed quantile-based trading strategies consist of three steps that are performed each day. In step (1), based on the median forecast ($\hat P^{50\%}$), the trader selects two hours -- one with the lowest price on a given day, denoted by $h_1$, and one with the highest price, denoted by $h_2$. However, if the battery is fully discharged on day $d^*$ ($B_{d^*}=0$), an additional unlimited bid will be placed at hour $h^* < h_2$. This requires not only choosing two hours where the minimum and maximum prices are most likely to be met, but also selecting hour $h^*$ corresponding to the additional bid, which is no longer a trivial task. To solve this problem, the linear optimization solver is used to select the optimal (i.e., most profitable) hours to place bids. The maximized function is equivalent to the following problem (the fractions $0.9$ and $\frac{1}{0.9}$ correspond to the battery efficiency):

\begin{equation}
	\max_{h_1, h_2, h^*} \left( 0.9~ \hat P^{50\%}_{d,h_2} - \frac{1}{0.9}\hat P^{50\%}_{d,h_1} - \frac{1}{0.9}\hat P^{50\%}_{d,h^*} \right), \quad \text{	subject to: $h^* < h_2$}.
\end{equation}
Similarly, if $B_{d^*} = 2$, the company will place an additional unlimited offer to sell energy before hour $h_2$. To select the transaction hours, the following problem must be solved:
\begin{equation}
	\max_{h_1, h_2, h^*} \left( 0.9~ \hat P^{50\%}_{d,h_2} - \frac{1}{0.9}\hat P^{50\%}_{d,h_1} + 0.9\hat P^{50\%}_{d,h^*} \right), \quad \text{	subject to: $h^* < h_1$}.
\end{equation}

In step (2), having chosen the optimal hours for placing bids, the trader has to make a decision about the level of the bid and the offer. First, the trader arbitrarily chooses the level $\alpha$ of the prediction interval (PI). The bid price is equal to the upper bound of the PI at hour $h_1$ i.e., $\hat U_{d,h_1}^{\alpha} = \hat P_{d,h_1}^{\frac{1+\alpha}{2}}$, while the offer price is set equal to the lower limit of the PI at hour $h_2$ i.e., $\hat L_{d,h_2}^{\alpha} = \hat P_{d,h_1}^{\frac{1-\alpha}{2}}$. The prices $\hat U_{d,h_1}^{\alpha}$ and $\hat L_{d,h_1}^{\alpha}$ are marked with black dots in Figure \ref{fig:stategy}. The additional unlimited bid or offer submitted on day $d^*$ the price-taker transactions. Therefore, it is not necessary to set the price level for hour $h^*$.

Finally, in step (3), the profit for day $d$ is calculated depending on the acceptance of the bid and offer:
\begin{itemize}
	\item If both the bid and the offer are accepted, the firm pays $P_{d,h_1}$ for $\frac{1}{0.9}$ MWh of energy bought at hour $h_1$ and receives $P_{d,h_2}$ for 0.9 MWh of energy sold at hour $h_2$. The daily profit is $0.9~P_{d,h_2} - \frac{1}{0.9}P_{d,h_1}$, and since the battery has been charged and discharged once, its state is unchanged ($B_{d+1} = B_{d}$).

	\item If only the offer is accepted, at hour $h_2$ the battery is discharged and 0.9 MWh of energy is sold on the day-ahead market. The daily profit is $0.9P_{d,h_2}$ and the battery state is reduced ($B_{d+1} = B_{d}-1$).

	\item Similarly, if only the bid is accepted, at hour $h_1$ the firm buys $\frac{1}{0.9}$ MWh of energy in the day-ahead market and charges the battery. The daily loss is $\frac{1}{0.9} P_{d,h_1}$, but the state of the battery is increased ($B_{d+1} = B_{d}+1$).
\end{itemize}

Additionally:
\begin{itemize}
	\item If $B_d = 0$, the trader additionally buys $\frac{1}{0.9}$ MWh of energy at hour $h^*$, so the daily profit decreases by $\frac{1}{0.9}P_{d,h^*}$ while the battery state increases.
	\item If $B_d = 2$, the battery owner additionally sells $0.9$ MWh of energy at hour $h^*$, so the daily profit increases by $0.9P_{d,h^*}$ while the battery state decreases.
\end{itemize}


\subsubsection{Unlimited-bids benchmark}
\label{sssec:benchmark:strategies}
If only the point forecast is available, a similar strategy can be proposed that does not require any knowledge of the price distribution.  In this case, the trader selects the hours with the lowest ($h_1$) and the highest ($h_2$) predicted prices (according to the point forecasts) each day. Unlimited (price taker) offers to sell energy at the highest price and to buy energy at the lowest price are then submitted. The daily profit is calculated taking into account the efficiency of the battery and is equal to $0.9P_{d,h_2} - \frac{1}{0.9}P_{d,h_1}$.

\section{Results}
\label{sec:results}

\subsection{Statistical measures}
\label{ssec:res:prob}

According to \citet{gne:raf:07}, the goal of evaluating probabilistic predictions should be to maximize sharpness subject to reliability. Reliability refers to the coverage, i.e. the percentage of real prices that fall within the prediction interval. Sharpness, on the other hand, describes the width of the interval. Therefore, prediction intervals should be as narrow as possible, but first they must provide reliable coverage. 
\subsubsection{Reliability}

Following this idea, the first step in evaluating probabilistic forecasts is to assess the reliability of the predictions. This can be done by defining a function that assigns a value of one whenever the actual price falls within the prediction interval (PI), and zero otherwise.
\begin{equation}\label{eq:coverage}
	I_{d.h}^{\alpha} = 
	\begin{cases}
		1, & \mbox{for } P_{d,h} \in [\hat L_{d,h}^{\alpha}, \hat U_{d,h}^{\alpha}],\\
		0, & \mbox{for } P_{d,h} \notin [\hat L_{d,h}^{\alpha}, \hat U_{d,h}^{\alpha}],
	\end{cases}
\end{equation} 
where $\hat L_{d,h}^{\alpha}$ is the lower bound and $\hat U_{d,h}^{\alpha}$ is the upper bound of the $\alpha$ prediction interval. In this paper, the PI coverage probability \citep[PICP;][]{now:wer:18} is reported. For the nominal coverage $\alpha$, the measure is defined as the percentage of times the real price is found within the $\alpha$ PI, averaged over all hours:
\begin{equation}
	\mbox{PICP}^{\alpha} = \frac{1}{24} \sum_{h=1}^{24}  \frac{\#(I_{d,h}^{\alpha} = 1:d=1,2,\ldots, D)}{D} \times 100\%,
\end{equation}
where $D = 2377$ is the total number of days in the out-of-sample period for probabilistic forecasts, see Section \ref{sec:data}. 

\begin{table}[b!]
	\caption{The prediction interval coverage probability (PICP) for $\alpha = 50\%, 70\%, \text{and } 90\%$. The values that are within $\pm 2.5\%$ of the nominal coverage are marked in green, while those that are beyond $\pm 5\%$ of the nominal coverage are marked in red. The value of PICP that is closest to the nominal coverage is marked in bold for each market separately.\\
		Note: **, * indicate significance at the 5\%, 1\% level for which at least half of hours (12 or more) pass the \cite{kup:95} test.}
	\label{tab:Cov}
	\scalebox{1}{
		\begin{tabular}{r|cc|cc|cc}
			\multicolumn{1}{c|}{}                       & \multicolumn{2}{c|}{50\%}                                                      & \multicolumn{2}{c|}{70\%}                                                      & \multicolumn{2}{c}{90\%}                                    \\
			\multicolumn{1}{c|}{\multirow{-2}{*}{PICP}} & EPEX                                  & OMIE                                  & EPEX                                  & OMIE                                  & EPEX                         & OMIE                         \\
			\midrule
			\midrule
			ARX-J                                      & 47.3                                  & 47.3                                  & 66.2                                  & 66.4                                  & 87.2                                  & 87.1                                  \\
			HS                                         & 47.3                                  & 47.1                                  & 66.3                                  & 66.4                                  & 87.0                                  & 86.9                                  \\
			CP                                         & \cellcolor[HTML]{C6E0B4}48.3**          & \cellcolor[HTML]{C6E0B4}48.1 **         & 67.2                                  & 67.3                                  & \cellcolor[HTML]{C6E0B4}87.8          & \cellcolor[HTML]{C6E0B4}87.6          \\
			QRA                                        & 45.0                                  & 45.3                                  & \cellcolor[HTML]{FFC7CE}63.6          & \cellcolor[HTML]{FFC7CE}63.9          & \cellcolor[HTML]{FFC7CE}84.1          & \cellcolor[HTML]{FFC7CE}84.0          \\
			QRM                                        & 45.9                                  & 46.8                                  & \cellcolor[HTML]{FFC7CE}64.9          & 65.9                                  & 85.5                                  & 85.9                                  \\
			QRF                                        & 46.8                                  & \cellcolor[HTML]{C6E0B4}47.6*          & 65.7                                  & 66.8                                  & 86.2                                  & 86.7                                  \\
			SQRA                                       & \cellcolor[HTML]{C6E0B4}48.4**          & \cellcolor[HTML]{C6E0B4}48.6**          & \cellcolor[HTML]{C6E0B4}67.7*          & \cellcolor[HTML]{C6E0B4}68.1*          & 86.7                                  & 87.0                                  \\
			SQRM                                       & \cellcolor[HTML]{C6E0B4}49.5**          & \cellcolor[HTML]{C6E0B4}\textbf{50.1}** & \cellcolor[HTML]{C6E0B4}68.3**          & \cellcolor[HTML]{C6E0B4}69.2**          & \cellcolor[HTML]{C6E0B4}87.6          & \cellcolor[HTML]{C6E0B4}88.0          \\
			SQRF                                       & \cellcolor[HTML]{C6E0B4}\textbf{50.4}** & \cellcolor[HTML]{C6E0B4}50.8**         & \cellcolor[HTML]{C6E0B4}\textbf{69.1}** & \cellcolor[HTML]{C6E0B4}\textbf{70.1}** & \cellcolor[HTML]{C6E0B4}\textbf{88.1} & \cellcolor[HTML]{C6E0B4}\textbf{88.6}*
		\end{tabular}
	}
\end{table}

To assess the accuracy of the forecasts, a comparison can be made with the nominal coverage, $\alpha$. The closer the value of the PICP is to $\alpha$, the more accurate the forecasts are in terms of reliability. In addition, to formally examine the reliability of the obtained prediction intervals (PIs), the \citet{kup:95} test has been employed for each of the 24 hourly series.

In Table \ref{tab:Cov}, the PICP measure is reported for three levels of nominal coverage ($\alpha=50\%, 70\%,$ and $90\%$). Asterisks are used to denote significance level for which at least half of hours (12 or more) pass the \cite{kup:95} test\footnote{Detailed results of \cite{kup:95} test are available in \ref{app:Kupiec}}.
Several conclusions can be drawn from the results:

\begin{itemize}\itemsep 0pt
	\item The SQRF model performs best in terms of prediction reliability. It produces PIs closest to the nominal coverage in five out of six cases. Only for the 50\% PI for the OMIE market is it the second best.
	
	\item All SQR Averaging variants provide reliable forecasts (according to Kupiec test) for the majority of hours for 50\% and 70\% PIs.
	
	\item The vast majority of methods struggle to provide reliable forecasts for the 90\% PI. Only SQRF passes the 1\% Kupiec test for the majority of hours for the OMIE market.
	
	\item In a pairwise comparison between QRA vs. SQRA, QRM vs. SQRM, and QRF vs. SQRF, the latter produced more reliable prediction intervals, in terms of PICP, for all considered cases.
	
		
	\item For all models except the SQR Averaging methods, all prediction intervals were too narrow. In some cases, they produced prediction intervals with empirical coverage lower than $\alpha$ by a significant margin (more than 6\%). Meanwhile, some SQR Averaging variants sometimes exceeded the nominal coverage for 50\% and 70\% prediction intervals.
\end{itemize}

\subsubsection{Sharpness}

The pinball score (PS) is a proper scoring rule \citep{pet:etal:22} and is widely used in the EPF literature \citep[for example,][]{mun:zie:20, kat:zie:21, nit:wer:23}. The measure is define as:

\begin{equation}\label{eq:Pinball}
	\mbox{PS} \left(\hat{P}^{q}_{d,h},P_{d,h},q \right) = 
	\begin{cases} (1-q) \left(\hat{P}^{q}_{d,h} - P_{d,h} \right) &\mbox{for } P_{d,h} < \hat{P}^{q}_{d,h}, \\ 
		q \left(P_{d,h} - \hat{P}^{q}_{d,h} \right) & \mbox{for } P_{d,h} \ge \hat{P}^{q}_{d,h},
	\end{cases}
\end{equation}
where $\hat{P}^{q}_{d,h}$ is the forecast of the price quantile of order ${q} \in(0,1)$ and $P_{d,h}$ is the observed price for day $d$ and hour $h$.

The pinball score can be averaged across quantiles and across load periods to obtain the aggregated pinball score (APS). If the grid of quantiles is dense, for example 99 percentiles, it can be used as an approximation of the continuous rank probability score \citep[CRPS;][]{lai:tam:07}. Here, in Table \ref{tab:Pinball}, the aggregated pinball score over all 99 percentiles and all hours of the out-of-sample test period (APS$_{99}$) is reported.

Since many financial and risk measures, such as Value at Risk (VaR) or Excess Return (ER), are based (primarily or exclusively) on extreme quantiles, Table \ref{tab:Pinball} reports the APS$_{10}$ aggregating the 5 lowest and 5 highest percentiles (10 in total) and all hours from the out-of-sample test period. Additionally to draw statistically significant conclusions, the \citet{gia:whi:06} test for conditional predictive ability (CPA) is performed with respect to SQRF model\footnote{Results of \citet{gia:whi:06} test for each pair of models is available in \ref{app:GW}}. Several conclusions can be drawn based on the results presented in the Table \ref{tab:Pinball}:

	\begin{itemize} \itemsep 0pt
		\item SQRF approach outperforms all other methods in terms of both APS$_{99}$ and APS$_{10}$. 
		\item Among the three variants considered for both QR and SQR Averaging models, the QRA and SQRA variants are consistently outperformed by the other two variants.
		\item Pairwise comparisons between QRA vs. SQRA, QRM vs. SQRM, and QRF vs. SQRF show that the latter models produce more accurate forecasts in all considered cases.
		\item Forecasts obtained with any benchmark (ARX-J, HS, CP) are inferior to QRF or SQRF in terms of sharpness.
		\item Surprisingly, CP is outperformed by both ARX-J and HS in terms of APS$_{99}$, which is inconsistent with the results of reliability measures.
	\end{itemize}

\begin{table}[tb]
	\caption{The Aggregate Pinball Score (APS) over all 99 percentiles (CRPS) and 10 extreme quantiles. The best score for each column is shown in bold. \\
		Note: **, * indicate significance at the 1\%, 5\% level of the \cite{gia:whi:06} test with respect to SQRF}
	\label{tab:Pinball}
	\scalebox{1}{
		\begin{tabular}{r|cc|cc}
			\multicolumn{1}{c|}{\multirow{2}{*}{}} & \multicolumn{2}{c|}{APS$_{99}$}        & \multicolumn{2}{c}{APS$_{10}$} \\
			\multicolumn{1}{c|}{}                     & EPEX           & OMIE           & EPEX            & OMIE            \\
			\midrule
			\midrule
			ARX-J                                    & 5.860**          & 3.717**          & 1.813**           & 1.187**           \\
			HS                                       & 5.856**          & 3.716**          & 1.794**          & 1.178**           \\
			CP                                       & 5.873**          & 3.723**          & 1.779**          & 1.168**           \\
			QRA                                      & 5.869**          & 3.727**          & 1.845**           & 1.214**           \\
			QRM                                      & 5.805**          & 3.700**          & 1.744**           & 1.166**           \\
			QRF                                      & 5.785**          & 3.681**          & 1.722**           & 1.139**           \\
			SQRA                                     & 5.844**          & 3.713*          & 1.801**           & 1.192**           \\
			SQRM                                     & 5.791**          & 3.691**          & 1.718**           & 1.147**           \\
			SQRF                                     & \textbf{5.782} & \textbf{3.679} & \textbf{1.705}  & \textbf{1.124} 
		\end{tabular}
	}
\end{table}


\subsection{Economic value}
No matter how accurate the forecasts are in terms of reliability or sharpness measures, their value is limited if the gain in accuracy cannot be converted into financial gains. In this section the results of the trading strategies described in Section \ref{ssec:strategies} are evaluated in terms of financial profits. The measure reported is the average amount of money the company can make from a single 1 MWh transaction. Therefore, the profit is defined as the total profit earned during the out-of-sample period divided by the sum of the total volume traded. Note that reported profits do not include transaction or investment costs.

\begin{figure}[tb]
	\centering
	
	\includegraphics[width=0.95\textwidth]{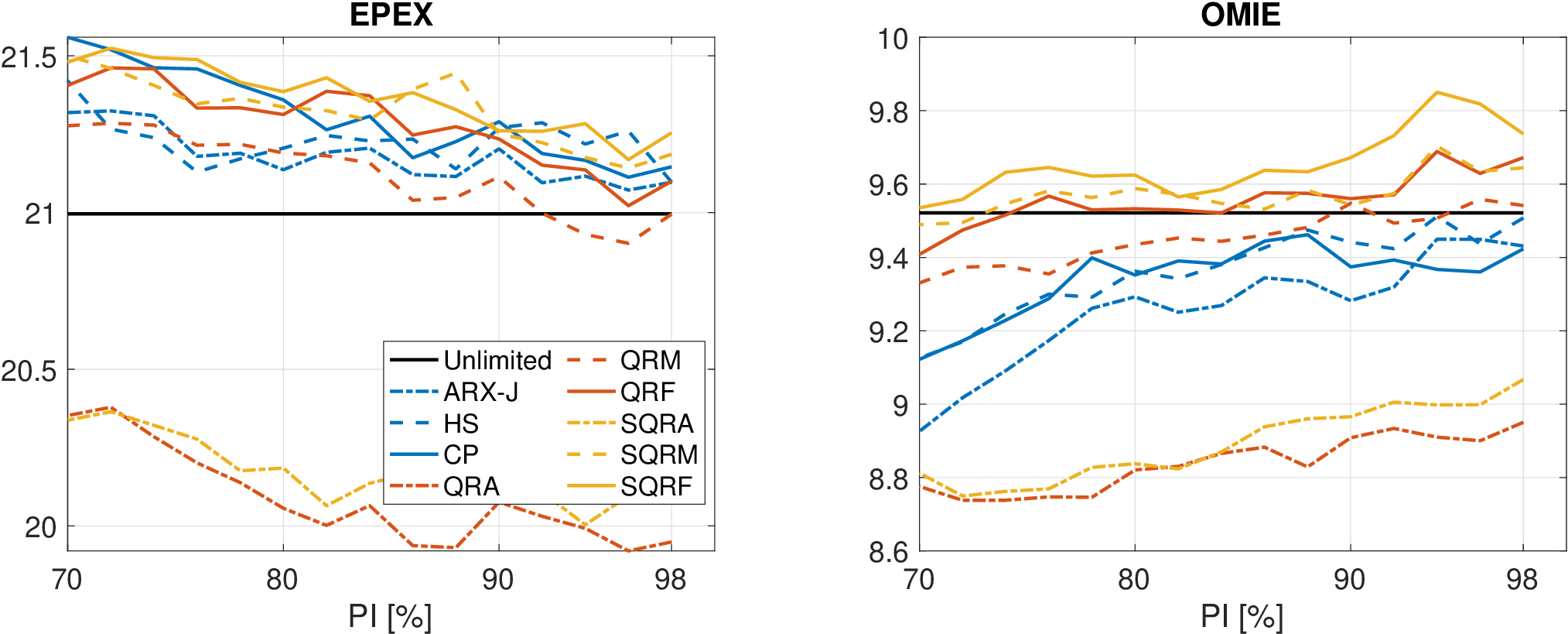}
	\caption{The profits per 1 MWh traded for the EPEX (left panel) and OMIE (right panel) markets obtained using benchmark and quantile-based strategies at different PI levels and all considered probabilistic forecasting models. The blue lines refer to the profits of the benchmark models, the red lines represent the class of QR Averaging models, and the yellow lines represent the profits of the SQR Averaging class of methods. The result of the unlimited bids strategy is plotted with solid black lines.}
	\label{fig:strategy}
\end{figure}

Figure \ref{fig:strategy} shows the profits (in EUR/MWh) obtained using the benchmark and quantile-based strategies for prediction intervals ranging from 70\% to 98\%. It can be seen that the quantile-based trading strategy shows significant improvements over the unlimited bidding strategy. In particular, using the 94\% prediction interval obtained with SQRF increases profits by up to 3.5\% for the OMIE market. Moreover, using the proposed trading strategy, market participants can obtain profits reaching more than 21.5 EUR per 1 MWh transaction, which is more than 43 EUR per charge-discharge cycle. Although the performance of the forecasting models varies across quantiles, especially in the EPEX market, the SQRF model generates the highest average profits. It clearly outperforms its competitors for OMIE and is among the top performers for the EPEX market. On the other hand, QRA and SQRA are consistently shown to be the worst forecasting models in terms of economic value for both markets, which is in line with the results presented in Tables \ref{tab:Cov} and \ref{tab:Pinball}.

\section{Conclusions}
\label{sec:conclusion}

The contribution of this paper is twofold. First, it introduces a new probabilistic forecasting technique called Smoothing Quantile Regression Averaging, which combines a well-performing load and price forecasting approach with kernel estimation to improve the reliability of the estimates. This approach produces probabilistic forecasts that significantly outperform a number of well-performing benchmarks. Second, the manuscript is one of the first in the literature to bridge the gap between computing probabilistic forecasts and trading based on such forecasts. More specifically, it introduces a trading strategy using probabilistic price forecasts and battery storage that can be effectively implemented in a day-ahead electricity market. 

Overall, the proposed SQR Averaging method offers significant potential for improving forecast accuracy. In particular the results of the empirical tests show that SQRF outperforms all the competitors in terms of both reliability and sharpness as well as economic value. The consistency of results, indicating the superior performance over other methods, suggests that SQR Averaging can be successfully applied to other markets and forecasting applications, which can be further explored in future research.

Future research should address the issue of bandwidth selection for the SQR Averaging method. In this paper, a simple rule of thumb of \citet{sco:92} is used. However, there are many other methods for selecting the optimal bandwidth, such as maximum likelihood cross-validation or the plug-in method. Despite the popularity of these alternative methods in the literature, their performance in this application remains an open research question. Therefore, further research is needed to determine the most appropriate bandwidth selection method for SQR-based models.

\section*{Acknowledgments}
This work was partially supported by the National Science Center (NCN, Poland) through MAESTRO grant No. 2018/30/A/HS4/00444. 

This paper has also benefited from conversations with the participants of the International Ruhr Energy Conference (INREC 2021), the Energy Finance Italia (EFI 2022), and the International Symposium on Forecasting (ISF 2022). 

\section*{Disclosure of interest}
No potential competing interest was reported by the author.

\section*{Data Availability Statement}
Raw data are openly available in ENTSO-E transparency platform at \url{https://transparency.entsoe.eu}. Derived data supporting the findings of this study are available from the corresponding author BU on request.

\bibliographystyle{elsarticle-harv}
\bibliography{references}
\newpage
\appendix

\section{Results of Kupiec test}
\label{app:Kupiec}
\begin{table}[H]
	\caption{The number of hours of the day (out of 24) for which the null hypothesis of the Kupiec test is not rejected at the 1\% (top) and 5\% (bottom) significance levels for $\alpha = 50\%, 70\%, \text{and } 90\%$ prediction intervals. The green color is used to indicate forecasts that pass the test for at least 12 hours. The highest value in each column is bolded.}
	\scalebox{1}{
		\begin{tabular}{r|cc|cc|cc}
			\multicolumn{1}{c|}{}                       & \multicolumn{2}{c|}{50\%}                                                      & \multicolumn{2}{c|}{70\%}                                                      & \multicolumn{2}{c}{90\%}                                    \\
			\multicolumn{1}{c|}{\multirow{-2}{*}{Kupiec}} & EPEX                                  & OMIE                                  & EPEX                                  & OMIE                                  & EPEX                         & OMIE                         \\
			\midrule
			\midrule
			ARX-J                                        & 11                                  & 10                                  & 2                                   & 2                                   & 0           & 0                                   \\
			HS                                           & 11                                  & 8                                   & 2                                   & 2                                   & 0           & 0                                   \\
			CP                                           & \cellcolor[HTML]{C6E0B4}\textbf{24} & \cellcolor[HTML]{C6E0B4}21          & 9                                   & 8                                   & 4           & 0                                   \\
			QRA                                          & 2                                   & 0                                   & 0                                   & 0                                   & 0           & 0                                   \\
			QRM                                          & 3                                   & 6                                   & 0                                   & 0                                   & 0           & 0                                   \\
			QRF                                          & 7                                   & \cellcolor[HTML]{C6E0B4}13          & 3                                   & 6                                   & 0           & 0                                   \\
			SQRA                                         & \cellcolor[HTML]{C6E0B4}18          & \cellcolor[HTML]{C6E0B4}22          & \cellcolor[HTML]{C6E0B4}14          & \cellcolor[HTML]{C6E0B4}18          & 0           & 1                                   \\
			SQRM                                         & \cellcolor[HTML]{C6E0B4}\textbf{24} & \cellcolor[HTML]{C6E0B4}\textbf{24} & \cellcolor[HTML]{C6E0B4}16          & \cellcolor[HTML]{C6E0B4}\textbf{24} & 1           & 6                                   \\
			SQRF                                         & \cellcolor[HTML]{C6E0B4}\textbf{24} & \cellcolor[HTML]{C6E0B4}\textbf{24} & \cellcolor[HTML]{C6E0B4}\textbf{23} & \cellcolor[HTML]{C6E0B4}\textbf{24} & \textbf{10} & \cellcolor[HTML]{C6E0B4}\textbf{17}
		\end{tabular}
	}
\vspace{10mm}

	\scalebox{1}{
		\begin{tabular}{r|cc|cc|cc}
			\multicolumn{1}{c|}{}                       & \multicolumn{2}{c|}{50\%}                                                      & \multicolumn{2}{c|}{70\%}                                                      & \multicolumn{2}{c}{90\%}                                    \\
			\multicolumn{1}{c|}{\multirow{-2}{*}{Kupiec}} & EPEX                                  & OMIE                                  & EPEX                                  & OMIE                                  & EPEX                         & OMIE                         \\
			\midrule
			\midrule
ARX-J & 4                                   & 4                                   & 1                                   & 0                                   & 0          & 0           \\
HS    & 4                                   & 2                                   & 0                                   & 0                                   & 0          & 0           \\
CP    & \cellcolor[HTML]{C6E0B4}\textbf{19} & \cellcolor[HTML]{C6E0B4}13          & 0                                   & 1                                   & 0          & 0           \\
QRA   & 0                                   & 0                                   & 0                                   & 0                                   & 0          & 0           \\
QRM   & 1                                   & 2                                   & 0                                   & 0                                   & 0          & 0           \\
QRF   & 6                                   & 6                                   & 2                                   & 2                                   & 0          & 0           \\
SQRA  & \cellcolor[HTML]{C6E0B4}15          & \cellcolor[HTML]{C6E0B4}18          & 7                                   & 10                                  & 0          & 0           \\
SQRM  & \cellcolor[HTML]{C6E0B4}\textbf{23} & \cellcolor[HTML]{C6E0B4}\textbf{24} & \cellcolor[HTML]{C6E0B4}14          & \cellcolor[HTML]{C6E0B4}\textbf{23} & 0          & 2           \\
SQRF  & \cellcolor[HTML]{C6E0B4}\textbf{21} & \cellcolor[HTML]{C6E0B4}\textbf{22} & \cellcolor[HTML]{C6E0B4}\textbf{18} & \cellcolor[HTML]{C6E0B4}\textbf{23} & \textbf{2} & \textbf{11}

		\end{tabular}
	}
\end{table}

\section{Results of pairwise CPA test}
\label{app:GW}
\begin{figure}[H]
	\centering 
	\includegraphics[width = 0.8\linewidth]{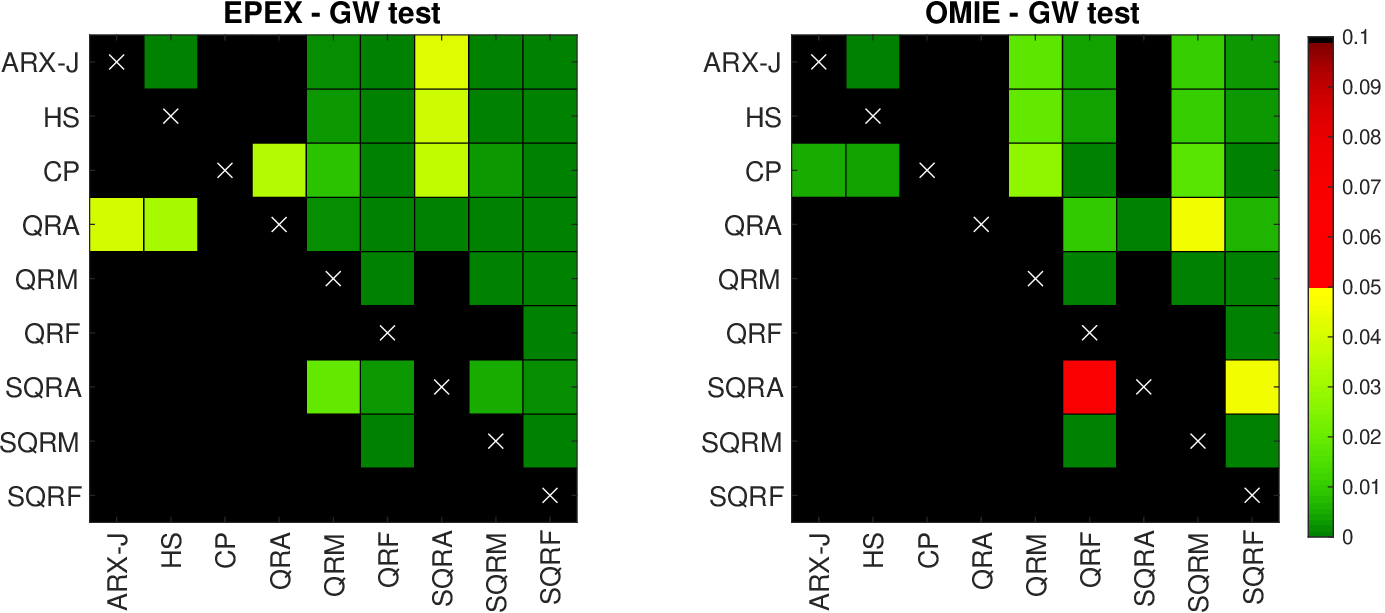}
	\caption{Results of the CPA test of \citet{gia:whi:06} for all considered probabilistic models.  Heat maps are used to illustrate the range of $p$-values -- the smaller they are ($\rightarrow$ dark green), the more significant the difference between the two forecasts (the model on the X-axis outperforms the model on the Y-axis).}
	\label{fig:GW}
\end{figure}

\end{document}